\documentclass[fleqn,12pt,twoside]{article}
\usepackage{espcrc1}

\usepackage{graphicx}

\newcommand{\AmS}{{\protect\the\textfont2
  A\kern-.1667em\lower.5ex\hbox{M}\kern-.125emS}}

\title{Spectral Change of Hadrons and Chiral Symmetry}

\author{T. Hatsuda\address{Physics Department, University of Tokyo,\\
       Tokyo 113-0033, Japan}}

\begin{document}

\maketitle

\begin{abstract}
After a brief summary of the QCD phase structure
 with light quarks, we discuss two recent developments
 on in-medium hadrons. First topic is  the 
 $\sigma$-meson which is a  fluctuation
  of the chiral order parameter $\bar{q}q$. Although $\sigma$
    is at best a broad resonance in the 
 vacuum, it may suffer a substantial red-shift and 
  show a characteristic spectral enhancement at the 
   2$m_{\pi}$ threshold at finite temperature and baryon density.
  Possible experimental signatures of this phenomenon are also discussed.
   Another topic is the first principle lattice QCD calculation
  of the hadronic spectral functions using 
   the maximum entropy method (MEM). The basic idea and a successful example 
    of MEM are presented.  Possible applications  of MEM to study the 
     in-medium hadrons in 
     lattice QCD simulations are discussed. 
      \end{abstract}

\section{Introduction}

One of the most intriguing phenomena in quantum chromo dynamics (QCD) is the 
dynamical breaking of chiral  symmetry.
This explains the existence of the pion and dictates
most of the low energy phenomena in hadron physics.
   The dynamical breaking of chiral symmetry is associated with the condensation of quark - anti-quark
pairs in the QCD vacuum, 
 $\langle \bar{q}q \rangle$, which is analogous to the condensation of 
 Cooper pairs 
 in the theory of  superconductivity \cite{NJL}.
 As the temperature ($T$) and/or the baryon density ($n_B$) increase,
 the QCD vacuum undergoes a phase transition to the chirally
 symmetric phase where $\langle \bar{q}q \rangle$ vanishes.
 Studying what are the 
  possible phase structure and how the phase transition takes place 
   are one of the main aims of the
  modern hadron physics \cite{HK94}.
 
 When we talk about the phase structure, we need to fix
 appropriate variables to define the phases.
  Such variables in QCD are current masses of light quarks ($m_{u,d,s}$),  $T$,
  and $n_B$.  Since the following relation holds, neglecting $m_{u,d}$ in
 the first approximation and studying the phase structure
 in the $m_s$-$T$-$n_B$ space   would give us
 a good insight into the real world:  
\begin{equation}
m_{u,d} \ll m_s \sim T_c \sim n_c^{1/3} \sim \Lambda_{QCD}.
\end{equation}
Here $T_c$ ($n_c$) are the critical temperature (density).
 In Fig.1, possible phase structures in the $m_s$-$T$ plane (with $n_B$=$0$)
 and $m_s$-$n_B$ plane (with $T$=$0$) are shown. 
 $m_s$=$0$ ($m_s$=$\infty $) corresponds to the limit of
  SU(3)$\otimes$SU(3) (SU(2)$\otimes$SU(2)) chiral symmetry.
 Recent lattice simulations show  $T_c \simeq 175 \ (155) $ MeV
  for 3 (2) flavors \cite{finiteT}. Furthermore, the first order
   transition for small $m_s$ turns into the  
     second order one for large $m_s$ at the tricritical
     point \cite{wilczek}.  
      The precise value of $m_s$ at the tricritical  point is
      not known yet.
             
       The phase transition at finite baryon density is not
       well understood  mostly because  the information from
       lattice QCD is poor. Naive analysis based on the
        bag model shows the first
       order transition \cite{heiselberg}, which implies the existence of a  
       mixed phase in the $m_s$-$n_B$ plane as shown in Fig.1 (right panel).
        However, the possibility of a smooth hadron-quark transition
        is not excluded.    The bag model analysis also suggests
        the strange quark matter as the true ground state of matter
     for sufficiently small $m_s$ \cite{mad}.
       The system at finite baryon density, which 
        is intrinsically quantum,  is obviously  richer in physics
       than that at finite $T$ and has 
       various phases such as the color superconductivity, meson condensations,
        and baryon superfluidity (see the reviews, \cite{RW00}).
        
\begin{figure}[t]
\begin{center}
\includegraphics[scale=0.45]{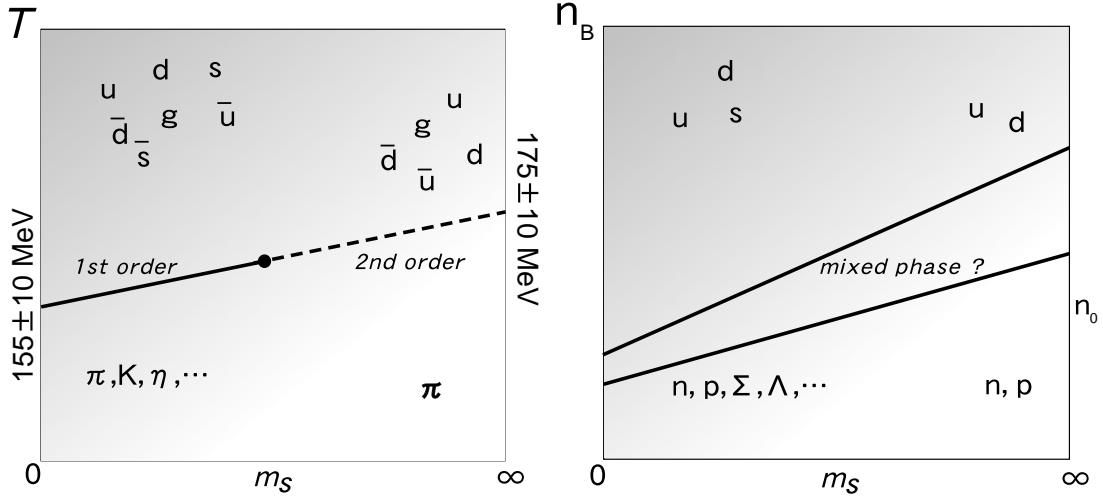}
 \end{center}
\caption{Possible phases in the $m_s$-$T$ plane at $n_B$=$0$ 
(left panel) and the $m_s$-$n_B$ plane at $T$=$0$. $m_{u,d}$=$0$ is assumed.
 Relevant degrees of freedom in each phase are also shown. $n_0 = 0.17{\rm fm}^{-3}$
  is the normal nuclear matter density.}
\label{fig1}
\end{figure}

Now, what are the observables associated with the chiral transition?
Model independent statement is that  hadronic correlations
in a same chiral multiplet (such as $S$ (scalar)$-$$P^a$ (pseudo-scalar) 
 and $V_{\mu}^a$ (vector)$-$$A_{\mu}^a$ (axial-vector))
  should be degenerate when $\langle \bar{q}q \rangle \rightarrow 0$, namely
\begin{eqnarray}
\langle S(x) S(y) \rangle \rightarrow \langle P^a(x) P^a(y) \rangle,
\ \ \ 
\langle A^a_{\mu}(x) A^b_{\nu}(y) \rangle
 \rightarrow \langle V^a_{\mu}(x) V^b_{\nu}(y) \rangle.
\end{eqnarray}
 Thermal susceptibilities for hadronic
  operators ${\cal O}$ defined in the Euclidian space,
\begin{equation}
\chi_{_H} = \int_0^{1/T} d\tau \int d^3x  \langle {\cal O}^{\dagger}(\tau, \vec{x})
{\cal O} (0,\vec{0})\rangle ,
\end{equation}
may be used to check such chiral degeneracy.
  Shown in Fig.2 is  the full QCD simulation of $\sqrt{1/\chi_{_H}}$
   on the lattice  
   with  2 light flavors \cite{karsch}.
  One can see the degeneracy 
  between the $\sigma$-channel ($I$=0, $J^P$=$0^+$)
   and the $\pi$-channel ($I$=1, $J^P$=$0^-$)
 above the critical point. There remains  a splitting
     between the $\delta$-channel ($I$=1, $J^P$=$0^+$)
 and the $\pi$-channel 
     above the critical point, which reflects
           the breaking of $U_A(1)$ symmetry.
 
   To make a direct connection with the experimental observables, however,
    one needs to go   further and calculate
  the Minkowski (real-time) correlation and  the hadronic spectral functions
    at finite $T$ and $n_B$ \cite{shu}. The aim of this talk is to show
   some recent attempts toward this goal.
   In Sec.2 and Sec.3, the spectral function in the $\sigma$-channel will be 
    discussed near 2$\pi$ threshold   where almost model-independent
     argument is possible.
 
 \begin{figure}[t]
\begin{center}
\includegraphics[scale=0.7]{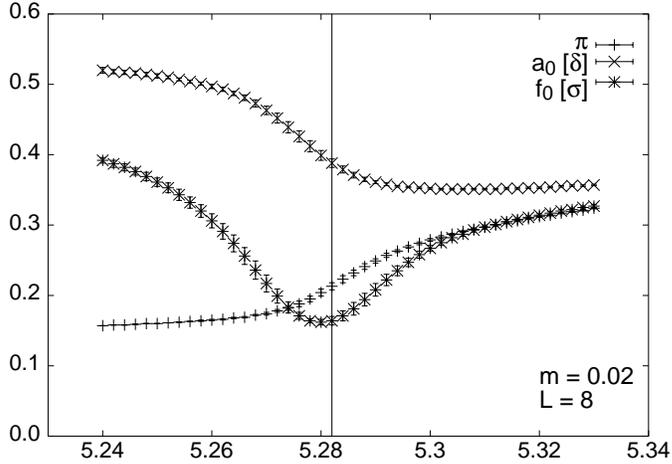}
\end{center}
\caption{Thermal susceptibilities in three different channels
 ($\pi$, $\sigma$, and $\delta$) for two-flavor QCD on the
   $8^3 \times 4$ lattice with  $m_{u,d} \ a = 0.02$ \cite{karsch}.
 The vertical (horizontal) axis denotes $\sqrt{1/\chi_{_H}}$ (the lattice coupling
  $\beta=6/g^2$). }
\label{fig2}
\end{figure}

 Experimentally, the resonances in the scalar and axial-vector channels 
have large width and  are  difficult to measure 
  with possible exception discussed in Sec.2,3.
 For this reason, in-medium neutral vector-mesons which couple directly to
 the virtual photon have been studied both theoretically and experimentally
  as a  probe of the chiral transition \cite{rw99}.
    Unfortunately, theoretical calculations of the spectral function
      in the vector channel
    on the basis of the effective Lagrangians are rather model dependent,
     and the first principle lattice QCD study has been called for.
    Recently, a new promising approach was proposed:
     it utilizes the maximum entropy method  
     for extracting the spectral functions from lattice QCD data. 
  We will discuss this new development in Sec.4.

\section{Spectral enhancement in hot plasma}
 
The fluctuation of the order parameter 
  becomes large as
 the system approaches to the critical point of a second order 
 or weak first order phase transition.
 In QCD,  the fluctuation of 
  the phase (amplitude) of the chiral order parameter 
 $\langle \bar{q}q \rangle $
  corresponds to  $\pi$ ($\sigma$).
    The vital role of such fluctuations in the 
  {\em dynamical} phenomena near the critical point
   at finite $T$ 
  was originally studied in \cite{HK85,rw94}. 
  
  It was suggested 
   that  the chiral restoration gives rise to a 
  softening (the red-shift) of the $\sigma$, which in turn  leads to
   a small  $\sigma$-width 
  due to the suppression of the phase-space of the decay
  $\sigma \rightarrow 2 \pi$  \cite{HK85}. Therefore, $\sigma$ may 
  appear as a narrow resonance at finite $T$
  although it is at best 
  a very broad resonance in the free space with a  width comparable to its
  mass \cite{pipi}.  (For the phenomenological applications of this  
 idea  in relation  to the relativistic heavy ion collisions, see 
 \cite{later}.)
 
  Further theoretical analysis   by taking  account the
  coupling $\sigma \leftrightarrow 2\pi$  shows that  
 (i) the spectral function  is the most relevant quantity
 for studying the true 
  nature of $\sigma$, and (ii)  the spectral function of  $\sigma$ 
 has a characteristic enhancement at the
  2$\pi$ threshold near  $T_c$ \cite{CH98}.   
     In Fig.3, shown are the 
     spectral functions $\rho_{\pi (\sigma)}$ in the $\pi$ ($\sigma$)-channel
      at finite $T$
 calculated in the $O(4)$ linear $\sigma$ model:
 Two characteristic features are 
 the broadening of the pion peak (Fig.3(A)) and the
    spectral concentration at  the 2$\pi$ threshold 
 ($\omega \simeq  2m_{\pi}$)  in the $\sigma$-channel (Fig.3(B)).
 The latter 
     may be measured by the $2\gamma$
    spectrum from the hot plasma created in the 
     relativistic heavy ion collisions \cite{CH98}.

 \begin{figure}[t]
 \begin{center}
\includegraphics[scale=0.5]{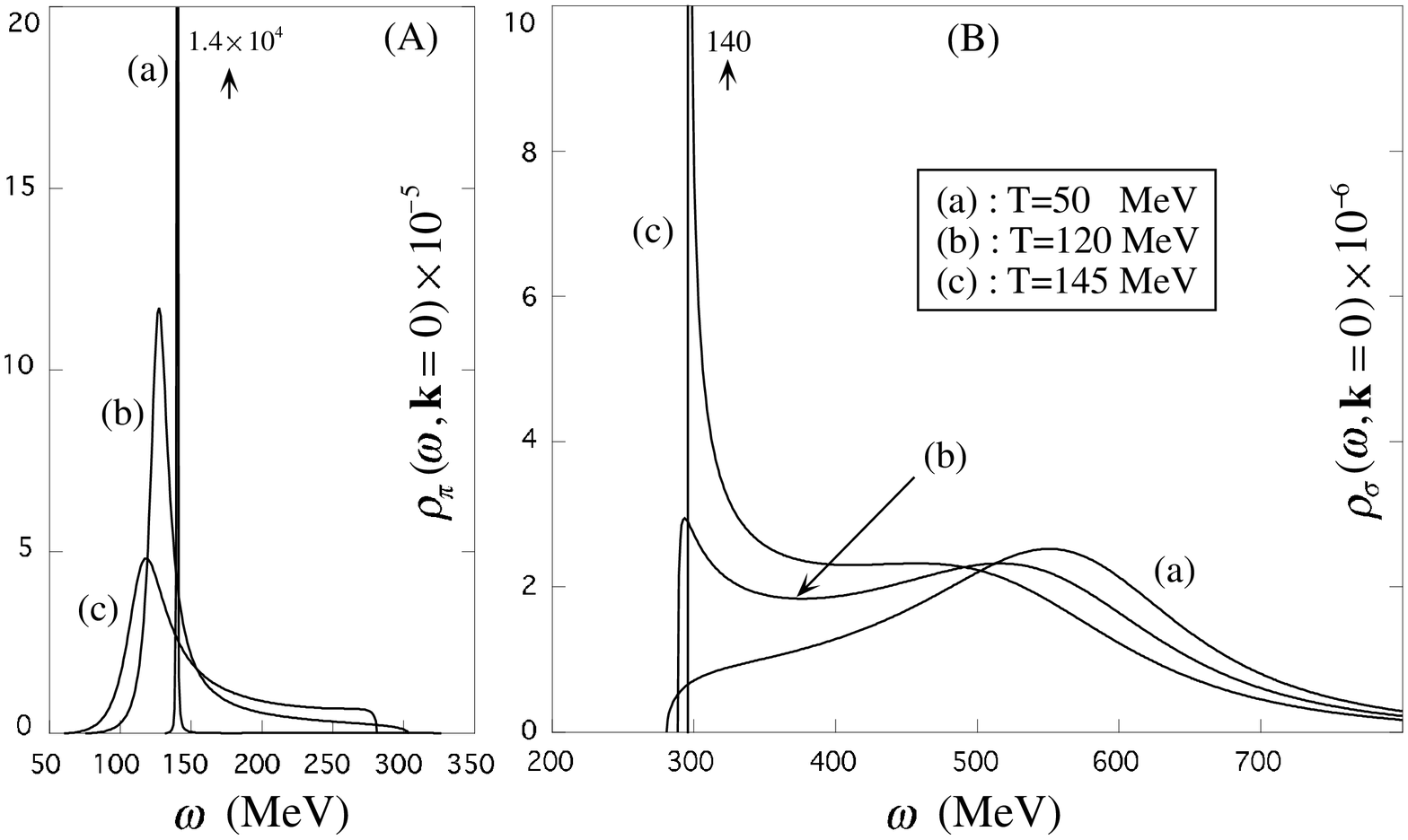}
\end{center}
 \caption{ Spectral functions in the $\pi$ channel (A) and in the
 $\sigma$ channel (B) for $T$=50, 120,  and 145 MeV. \cite{CH98}
 $\rho_{\sigma}(\omega)$ in (B) shows only a broad bump at low $T$ (a), while
 the spectral concentration is developed as $T$ increases,
  (a)$\rightarrow$(b)$\rightarrow$(c).}
  \label{fig3}
\end{figure}

\section{Spectral enhancement  in nuclear matter}
 
  $\langle \bar{q}q \rangle$
 at finite baryon density obeys  an exact theorem in QCD  \cite{DL}:
 \begin{eqnarray}
 \label{cond-rho}
{\langle \bar{q} q \rangle \over
 \langle \bar{q} q \rangle_{0} }
= 1 - {n_B \over f_{\pi}^2 m_{\pi}^2 } \left[ \Sigma_{\pi N}
 + \hat{m} {d \over d\hat{m}} \left( {E(n_B)  \over A } \right) \right] ,
 \end{eqnarray}
where $\Sigma_{\pi N}= 45 \pm 10 $ MeV is the pion-nucleon sigma term
 and $E(n_B)/A$ is the nuclear binding energy per particle with $\hat{m}
 =(m_u + m_d)/2$. 
$\langle \bar{q} q \rangle_{0}$
 denotes the chiral condensate in the vacuum.
The  density-expansion of the right hand side of (\ref{cond-rho})
 gives a reduction of almost 35 \% of $\langle \bar{q} q \rangle $
 already at the nuclear matter density $n_0 = 0.17
 $fm$^{-3}$.

The near-threshold enhancement discussed in Sec.3
  has been also studied at finite baryon density \cite{HKS}.
 Because of the decrease of the chiral condensate discussed above,
   the spectral function in the $\sigma$-channel, $\rho_{\sigma}$,
   has  a  following generic behavior at densities  not far
    from $n_0$;
        \begin{eqnarray}
\rho_{\sigma} (\omega \simeq  2 m_{\pi}) 
  \propto {\theta(\omega - 2 m_{\pi}) 
 \over \sqrt{1-{4m_{\pi}^2 \over \omega^2}}},
 \end{eqnarray}
which  shows a spectral concentration at the $2\pi$
threshold. 

 \begin{figure}[t]
\begin{center}
\includegraphics[scale=0.7]{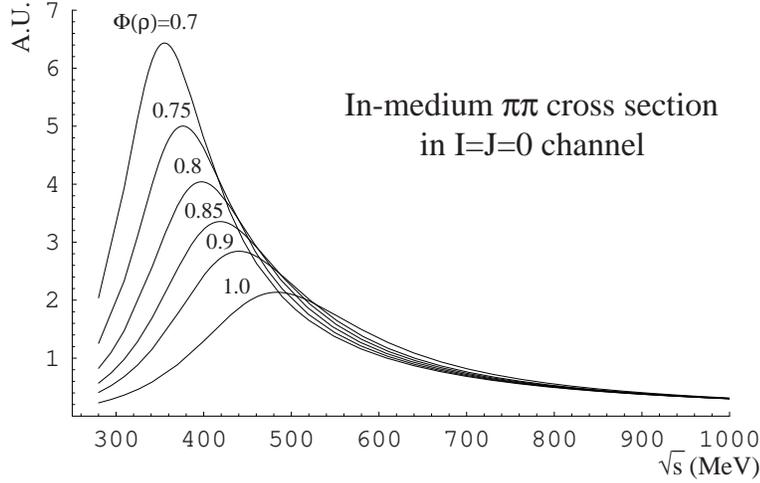}
\end{center}
\caption{In-medium $\pi\pi$ cross section in the $I$=$J$=$0$ channel
 as a function of the c.m. energy $\sqrt{s}$
  for several different values of the condensate ratio
  $\Phi=\langle \sigma \rangle /\langle \sigma \rangle_0$ in
   nuclear matter \cite{JHK}. }
\label{fig3}
\end{figure}

Recently  CHAOS collaboration  \cite{CHAOS} reported the data on the
$\pi^{+}\pi^{\pm}$
invariant mass distribution $M^A_{\pi^{+}\pi^{\pm}}$ in the
 reaction $A(\pi^+, \pi^{+}\pi^{\pm})A'$ with the 
 mass number $A$ ranging
 from 2 to 208: They observed that
the   yield for  $M^A_{\pi^{+}\pi^{-}}$ 
 near the 2$\pi$ threshold is close to zero 
for $A=2$, but increases dramatically with increasing $A$. They
identified that the $\pi^{+}\pi^{-}$ pairs in this range of
 $M^A_{\pi^{+}\pi^{-}}$ is in the $I$=$J$=$0$ state.
This experiment was
 originally motivated by a possibility of strong
 $\pi\pi$ correlations in nuclear matter \cite{GSI}.
  However,  the state-of-the-art
  calculations using the nonlinear chiral
  Lagrangian together with  $\pi N \Delta$ many-body
  dynamics do not reproduce the cross sections in the $I$=$0$ and
  $I$=$2$ channels simultaneously  \cite{WO};
 the final state interactions of the emitted two pions in nuclei
  give rise to a slight enhancement of the cross section in the
  $I$=$0$ channel, but is
  not sufficient to reproduce the experimental  data.
 This indicates that
 an additional mechanism such as the partial restoration of chiral
 symmetry   may be
 relevant for explaining the data \cite{HKS,SHUCK}.

To make a close connection between the 
 idea of the spectral enhancement and the CHAOS data,
  the in-medium $\pi$-$\pi$ cross section has been calculated in 
   the linear and non-linear $\sigma$ models \cite{JHK}.
   It was shown that, in  both cases, substantial enhancement of the 
   $\pi$-$\pi$ correlation in the $I$=$J$=$0$ channel near the threshold can be seen 
   due to the partial restoration of chiral symmetry.
    Furthermore, an effective 4$\pi$-$N$-$N$ vertex responsible
     for the enhancement in the non-linear chiral Lagrangian is identified.
       In Fig.4, the in-medium $\pi$-$\pi$ cross section
   in the $\sigma$-channel
  calculated in the O(4) linear $\sigma$ model is shown. 
  
  Theoretically, it is 
  of great importance to make
 an extensive analysis of the $\pi$-$\pi$ interaction in heavy nuclei
 with  the effect of partial chiral restoration to understand the CHAOS data.
Experimentally, 
it is definitely necessary to confirm the CHAOS result not only in the same
 reaction  but also in  other reactions.
  Measuring $\sigma \rightarrow 2 \pi^0 \rightarrow
  4\gamma$ is one of the clean 
   experiments, since it is free from the $\rho$ meson background
  inherent in the $\pi^+\pi^-$ measurement.
 Measuring $\sigma \rightarrow 2 \gamma$
  is interesting because of the small final state
 interactions. Dilepton detection  
 through the scalar-vector mixing in matter, $\sigma \to \gamma^* \to
 e^+ e^-$  and 
 the formation of the $\sigma$ mesic nuclei through the 
    nuclear reactions such as (d, t) are 
    other possible experiments  \cite{kuni95,HKS}.

\section{Spectral functions from lattice QCD }

 The
 lattice QCD simulations have remarkable progress in recent years for
 calculating the properties of hadrons as well as the 
 properties of QCD phase transition \cite{lattice}.
 In particular, masses of light mesons and baryons
  in the quenched QCD simulation   
  agree within 5-10 \% with
 the experimental spectra \cite{kanaya}. However, the lattice QCD
 had difficulties in accessing the dynamical quantities 
 in the Minkowski space, because
 measurements on the lattice can only be carried out
 for discrete points in imaginary time.
 The analytic continuation from the imaginary time to the
 real time using the noisy lattice data 
 is highly non-trivial  and is even classified as an ill-posed
 problem. 
 
  Recently a new approach  to
 extract spectral functions of hadrons from lattice
 QCD data by using the 
 maximum entropy method (MEM) was started  \cite{nah}.
 MEM  has been successfully applied for 
 similar problems in image reconstruction in crystallography and 
  astrophysics, and in 
  quantum Monte Carlo simulations in
 condensed matter physics \cite{physrep}.
 There are three important aspects of MEM: (i) it does not require a priori
assumptions or parametrizations of SPFs, (ii) for given data, 
a unique solution 
is obtained if it exists, and (iii)
the statistical significance of the solution can be quantitatively analyzed.

\subsection{Basic idea of MEM}

The Euclidean correlation function $D(\tau)$ of an 
operator ${\cal O}(\tau,\vec{x})$  and its spectral decomposition
at zero three-momentum read
\begin{eqnarray}
D(\tau > 0) = \int 
\langle {\cal O}^{\dagger}(\tau,\vec{x}){\cal O}(0,\vec{0})\rangle d^3
x\label{KA}  \equiv \int_{0}^{\infty} \!\! K(\tau,\omega) A(\omega ) d\omega,
\end{eqnarray}
where  $\omega$ is a real frequency, and
$A(\omega)$ is the spectral function 
(or sometimes called the {\em image}),
which is positive semi-definite. $K(\tau,\omega)$ is a known integral kernel
 (it reduces to Laplace kernel $e^{-\tau \omega}$ at zero $T$.)

Monte Carlo simulation provides  $D(\tau_i)$ 
on  the discrete set of temporal points $0 \le \tau_i /a \le N_\tau$.
From this  data with statistical noise, we need to reconstruct the 
spectral function $A(\omega)$ with continuous variable $\omega$. 
 This is a
typical ill-posed problem, where the number of data is much smaller than
the number of degrees of freedom to be reconstructed.
This makes the standard 
likelihood analysis and its variants
inapplicable  unless strong assumptions
on the spectral shape are made \cite{others}.
MEM is a method to circumvent this difficulty
 through Bayesian  statistical inference of the most probable {\em image}
 together with its reliability.

 Let's start with  the Bayes' theorem:
 $P[X|Y] = P[Y|X]P[X]/P[Y]$,
 where $P[X|Y]$ is the conditional probability of $X$ given $Y$.
 The most probable image 
 $A(\omega )$ for given lattice data $D$ is obtained by 
 maximizing the conditional probability 
 \begin{equation}\label{bayes_latt}
P[A|D] \propto P[D|A]P[A] .
\end{equation}
 The reliability of the obtained result
 can be checked by the second variation of $P[A|D]$ with 
 respect to $A(\omega)$.
 
$P[D|A]$ in (\ref{bayes_latt}) is   a standard $\chi^2$, namely 
$P[D|AH]= Z_L^{-1} \exp (-L)$ with a likelihood function 
 $L  = (1/2) \sum_{i,j}
(D(\tau_i)-D^A(\tau_i)) \,  C^{-1}_{ij} \, 
(D(\tau_j)-D^A(\tau_j))$ and a normalization factor $Z_L$.
$D(\tau_i )$ is the lattice data averaged over gauge configurations 
and $D^A(\tau_i )$ is the correlation function obtained by given
 $A$. $C$ is a covariance matrix of the data.
  
  $P[D|A]$ (the prior probability) in (\ref{bayes_latt}),
   can be written
with parameters $\alpha$ and $m$ as
$P[A]= Z_S^{-1} \exp (\alpha S)$ by the combinatorial argument.
 Here
$S$ is the generalized information entropy,
\begin{eqnarray}
S = \int_0^{\infty} \left [ A(\omega ) - m(\omega )
- A(\omega)\log \left ( \frac{A(\omega)}{m(\omega )} \right ) \right ]
d\omega , 
\end{eqnarray}
with $Z_S$ being a normalization factor.
$\alpha$ is a real and positive parameter and 
$m(\omega )$ is a real and positive function called the default model.
 
The output image $A_{out}$ is
given by a weighted average over $A$ and  $\alpha$:
\begin{eqnarray}
A_{out}(\omega)    
 =  \int A_{\alpha}(\omega) \   P[\alpha|Dm]  \ d\alpha ,
\label{final}
\end{eqnarray}
where $A_{\alpha}(\omega)$ is obtained by
minimizing  the "free-energy" $F \equiv L - \alpha S$. 
$\alpha$ dictates the relative weight of the
entropy  $S$ (which tends to fit $A$ to the default model $m$)
and the likelihood function $L$ (which tends to fit $A$ to the
lattice data). Note that 
$\alpha$ appears only in the intermediate
step and is  integrated out in the final result.

One can   prove that
 the solution of $\delta F=0$ is unique if it exits.
  The error analysis
  of the reconstructed image can be studied by evaluating
 the second derivative of the free energy
  $\delta^2 F /\delta A(\omega) \delta A(\omega') $.
  The default model $m$ is chosen such that the error becomes minimum.

\subsection{MEM with lattice data}

 In Fig.5, shown is an example of the spectral function
  in the pion and the rho meson channels at $T=0$ extracted from the 
  quenched lattice QCD data \cite{nah}. The lattice size is $20^3 \times 24$
with $\beta =6.0$, which corresponds to 
$ a = 0.0847$ fm ($a^{-1} = 2.33$ GeV),
 and the
spatial size of the lattice $L_s a = 1.69 $ fm.
Hopping parameters are chosen to be
$\kappa =$ 0.153, 0.1545, and 0.1557 with $N_{conf}=161$
for each $\kappa$.
For the quark propagator, the Dirichlet (periodic)
boundary condition
is employed for the temporal (spatial) direction.
We use data at $1 \le \tau_i/a  \le 12  $.

\begin{figure}[htb]
\begin{center}
\includegraphics[scale=0.8]{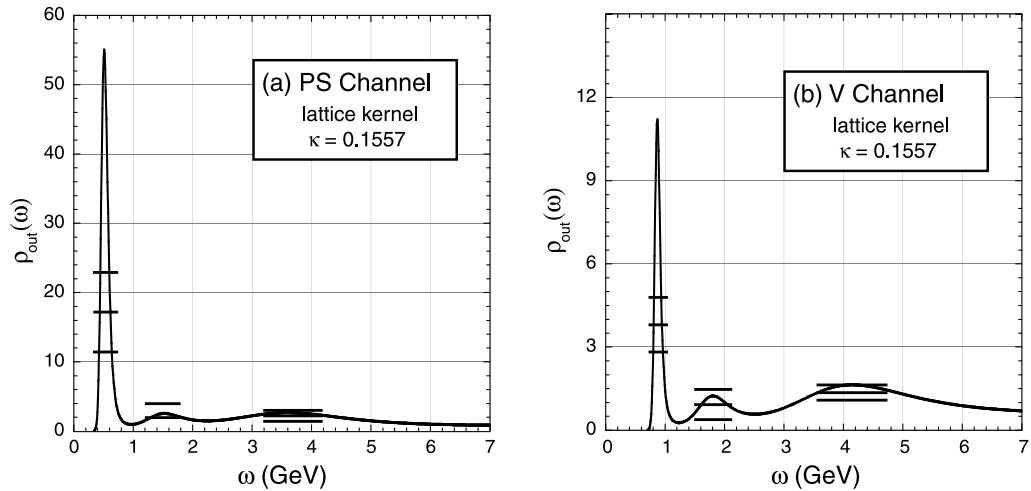}
\end{center}
\caption{Spectral functions $\rho_{out}(\omega) 
 \equiv A_{out}(\omega) /\omega^2$
 obtained by MEM using  the quenched  lattice QCD 
 data. The lattice size is 20$^3$$\times$24 with $a=0.0847$ fm.
 12 data points in the temporal direction ($1 \le \tau_i /a \le 12$) are
  used for the MEM analysis. (a) is for the pion channel and (b) is 
  for the rho-meson channel. The figures are taken from \cite{nah}.}
\label{fig5}
\end{figure}

The obtained images have the low-energy peaks corresponding to
$\pi$ and $\rho$, and the broad structure in the high-energy
region.  
 The mass of the $\rho$-meson in the chiral limit
extracted from the peak in Fig.5 is consistent with that
 determined by the asymptotic behavior of $D(\tau)$.
 Although the maximum value of the
fitting range $\tau_{max}/a =12$ 
marginally covers the asymptotic limit in $\tau$, we 
can extract reasonable masses for $\pi$ and $\rho$.
The width of $\pi$ and $\rho$  in Fig.5
is an artifact due to the statistical errors of the
lattice data. In fact, in the quenched approximation, there is no room 
for the $\rho$-meson to decay into two pions.

As for the second peaks,
the error analysis shows that
their spectral ``shape" does not have much  statistical
significance, although the existence of the
non-vanishing spectral strength is significant.
Under this reservation, we fit the position of the 
second peaks and made linear
extrapolation to the chiral limit and obtain  
  results consistent with experimental data on $\pi'$ and $\rho'$.
 
The error in Fig.5 indicates the uncertainty of the spectrum averaged over
 the interval in the frequency space.
  Better data with smaller $a$ and larger lattice volume will
 be helpful for obtaining spectral functions with smaller errors.

\subsection{The future of MEM}

  MEM introduces 
 a new way of extracting physical
  information from the lattice QCD
  data. It is now time to study how hadrons are
   modified in the medium using this new approach.
  The long-standing problem of in-medium spectral functions of
        vector mesons
         ($\rho$, $\omega$, $\phi$, $J/\psi$, $\Upsilon$, $\cdots$, etc) 
        and scalar/pseudo-scalar mesons ($\sigma$, $\pi$, $\cdots$, etc)
        can be studied using MEM combined with finite $T$ lattice simulations.
        The in-medium behavior of
        the light vector mesons \cite{pisa} and
        scalar mesons \cite{HK85} is intimately
        related to the chiral
        restoration in hot/dense matter, while that of the 
        heavy vector mesons 
        is related to deconfinement \cite{HM}. Simulations with 
        an anisotropic lattice is necessary for this purpose
        to have enough
        data points in the temporal direction at 
        finite $T$, which is now under way \cite{now}. 
        (See also an attempt
        on the basis of NRQCD simulation \cite{ukqcd_ohio}.)
        It is interesting to study
        the spectral functions with finite three-momentum ${\bf k}$
        for treating moving-hadrons in the medium.
   
  Non-perturbative  
        collective modes above the critical temperature of 
        the QCD phase transition speculated in  \cite{HK85,DeTar,thoma}  
        may be studied efficiently by using MEM, 
        since the method does not
        require any specific ans\"{a}tze for the spectral shape.
        Correlations  in the diquark channels in the vacuum
        and in medium are 
        interesting to be explored in relation to the 
        $q$-$q$ correlation
        in baryon spectroscopy and to color superconductivity
        at high $n_B$ \cite{wk}.

\section{Summary}

Although the light $\sigma$-meson  does not show up 
 clearly in the free space because of its large  width due to 
the strong coupling with two pions,
 it may appear as a soft and narrow
 collective mode in the hadronic medium 
when the chiral symmetry is (partially) restored.
 A characteristic signal is the enhancement of the spectral function
 and the $\pi$-$\pi$ cross section near the  $2\pi$ threshold. They
 could be observed in the hadronic
 reactions with heavy nuclear targets as well as in the heavy
 ion collisions through the detection of pion pairs and photon pairs.
   Finding such signal provides us with a 
 better understanding of the non-perturbative structure of the 
 QCD vacuum and its quantum fluctuations.
 
 The maximum entropy method (MEM) opens the door to
  extract the spectral functions from the 
   first principle lattice QCD data without
 making a priori assumptions on the spectral shape.
 The method has been tested at $T$=$0$ with reasonable success.
  The uniqueness of the solution 
 and the quantitative error analysis make MEM
 superior to any other approaches adopted previously.
 The method has a promising future for analyzing
   in-medium properties of hadrons and its relation to
   chiral structure of matter.

\end{document}